\documentclass[12pt]{article}
\usepackage{epsfig}
\usepackage{bm}
\usepackage{amsmath}
\usepackage{amssymb}

\begin{document}
\parskip 0.3cm
\begin{titlepage}
\begin{centering}
{\large \bf NLO Semi-inclusive Drell-Yan cross-section in Quantum 
ChromoDynamics as a Factorization Analyzer } 

 \vspace{.9cm}
{\bf Federico Alberto Ceccopieri}\\
\vspace{.2cm}
and\\
\vspace{.2cm}
{\bf Luca Trentadue}
\vspace{0.7cm}
\\ {\it Dipartimento di Fisica, Universit\'a di Parma,\\
INFN Gruppo Collegato di Parma, Viale delle Scienze, Campus Sud,\\ 
43100 Parma, Italy}\\
\vspace{1cm}
{ \bf Abstract}
\vspace{0.5cm}
\\ 
\bigskip
\end{centering}
{\small \noindent We evaluate in perturbative QCD 
the semi-inclusive Drell-Yan cross-section for the production 
of a single hadron accompaining the lepton pair. 
We demonstrate to one loop level 
a collinear factorization formula within the fracture functions approach.
We propose such a process as a factorization analyzer in hadronic collisions.
Phenomenological implications at the hadron colliders are briefly discussed.} 

\vspace{7cm}
%\vfill
\begin{flushleft}
{\small electronic address: federico.alberto.ceccopieri@cern.ch\\
electronic address: luca.trentadue@cern.ch \hfill}
\end{flushleft}

\end{titlepage}
\vfill\eject

\noindent{\large \bf Introduction}\\

\noindent
The Drell-Yan \cite{DY} process has unique features among high energy hadronic reactions.
The measurament of the relative cross-sections represents in fact a basic normalization in the LHC physics program, 
in particular by proving the universality of parton distribution functions as measured 
in Deep Inelastic Scattering and by possibly pinning down non-standard evolution
in the initial state parton cascade at small values of Bjorken variable $x$.

Universality of parton distribution functions follows directly from the factorization theorem for the 
Drell-Yan process. Despite some initial controversy on such a factorization in the early '80,  
a series of papers~\cite{soft_fact} supported this hypothesis with specific 
model calculations and then a factorization proof was finally given in Refs.~\cite{CSS,Bodwin}.
It was shown in particular that, when diagrams with soft 
gluon exchanges between active and spectators 
partons are summed over all possible cuts, 
soft gluon contributions decouple from the short distance cross-section, 
convalidating factorization at parton level as firstly conjectured by Drell and Yan. 
All the complications in proving the factorization can be related to the presence of 
two hadrons within the initial state. 
    
The Drell-Yan process has furthermore two more appreciable features 
with respect to other short-distance hadron-induced cross-sections:  
a) the perturbative scale can be accurately reconstructed by measuring the invariant mass $Q^2$ of the lepton pair
and b) the final state is free from QCD-corrections 
thus providing a clean tool for the study of initial state radiation pattern.

Given these properties, it may become natural to investigate what happens if an additional 
hadron, accompaining the lepton pair, is identified in the final state:
\begin{equation}
\label{process:SIDY}
P_1 + P_2 \rightarrow \gamma^* + h + X\,,
\end{equation} 
where $P_1$ and $P_2$ denote the incoming hadrons, 
$h$ is the identified hadron in the final state, 
$\gamma^*$ is the virtual (eventually electroweak) boson and $X$ the unobserved part of
the final state.
The idea of using the Drell-Yan process as a \textsl{perturbative} trigger 
were first investigated in Ref.~\cite{DeGrand} and will be fully exploited also at the LHC~\cite{CMS_TOTEM_TDR,joint}.
If the invariant mass $Q^2$ of the lepton pair is large enough so that perturbation theory applies, 
following the arguments of Ref.~\cite{DeTar}, the factorization property of the 
cross-section for the process in eq.~(\ref{process:SIDY}), should depend on the region of phase space 
in which the final hadron $h$ is detected. In particular if $h$ is produced at sufficiently high 
transverse momentum, $p_{h\perp}^2$, then in such phase space regions 
(we refer to them as to \textsl{central}) standard perturbative QCD technique 
should be applicable, see for istance Ref.~\cite{STAR}. If $h$ has instead a low $p_{h\perp}^2$ 
and thus is detected in the so-called \textsl{target} fragmentation region, 
arguments against factorization have been already given in Refs.~\cite{CFS,BS,Fact_M_soft}.

It is therefore highly desiderable to have a standard perturbative framework 
which can be used as a \textsl{''factorization analyzer''} in both region of phase space.  
The first step is thus to provide a parton model formula which accounts for the
production of an additional hadron. In the inclusive case the cross-section for the production of a lepton pair 
of mass $Q^2$ in the collision of two hadrons, of momenta $P_1$ and $P_2$, 
can be written as~\cite{DY}
\begin{equation}
\label{iDY}
\frac{d\sigma^{DY}(\tau)}{dQ^2}=\frac{4\pi\alpha^2}{9 S Q^2}
\int \frac{dx_1}{x_1} \int \frac{dx_2}{x_2}
\sum_q e_q^2 \Big[ f_q(x_1) f_{\bar{q}}(x_2)+(x_1 \leftrightarrow x_2) \Big] 
\delta\Big( 1-\frac{\tau}{x_1 x_2}\Big)\,, 
\end{equation}
with $S$ the hadronic center of mass energy, $S=(P_1+P_2)^2$, and $\tau=Q^2/S$ as in Ref.~\cite{DYNLO}.
The sum runs over all quarks and antiquarks
flavour but not on gluons which can not directly couple to electroweak bosons.
The parton distribution functions $f_q(x)$ depend on the fractional momentum of the 
parton entering the hard scattering. In particular 
no scale dependence is indicated as appropriate in the n\"aive parton model formula. 
In such an approach to the semi-inclusive Drell-Yan 
process of eq.~(\ref{process:SIDY}),
where QCD higher order corrections are absent,  
we assume that the final state hadron is "non-perturbatively" produced in the target 
fragmentation region of $P_1$ ($\mathcal{R}_{T_1}$) or $P_2$ ($\mathcal{R}_{T_2}$)
by means of a fracture function $M^i_{h/P}(x,z)$. 
These distributions give the conditional probability of finding 
a parton $i$ with a fractional momentum $x$ while an hadron $h$, with fractional momentum $z$ of 
the incoming hadron momentum $P$,
is detected in the target fragmentation region of $P$, see Ref.~\cite{Trentadue_Veneziano}.
The collinear and soft factorization of these distributions in semi-inclusive DIS 
has been proven respectively in Refs.~\cite{Fact_M_coll,Fact_M_soft} and by an explicit 
$\mathcal{O}(\alpha_s)$ QCD calculation in Ref.~\cite{Graudenz}.
Supported by these results, continous efforts have been devoted to the extraction
diffractive parton distributions from HERA data.
A combined analysis of both the diffractive and leading proton DIS data   
in terms of fracture functions has been presented in Ref.~\cite{DeFSassot}
while for more recent analysis we refer to Refs.~\cite{DPDF}.
Since no perturbative emissions are allowed in a parton model 
formula, we assume that "bare" fracture functions describe hadron production  
in the target fragmentation region 
$\mathcal{R}_{T_1}$ of $P_1$ if $\theta_{\makebox{\tiny{cm}}}=0$ and in $\mathcal{R}_{T_2}$ of $P_2$,
if $\theta_{\makebox{\tiny{cm}}}=\pi$,
where $\theta_{\makebox{\tiny{cm}}}$ is the relative angle between $h$ and $P_1$ in the 
hadronic center of mass frame.
Phase space separation between target and central region is indeed unphysical and 
in this particular case also frame dependent. However this choice can be shown to be the more suitable     
in order to prove the factorization of the collinear singularities of the cross-sections.   
In the following we will consider the next-to-simple differential cross-sections 
for producing a lepton pair of invariant mass $Q^2\gg \Lambda_{QCD}^2$, accompained by 
an additional hadron $h$ with fractional energy $z=2E_h/\sqrt{S}$ 
(defined in the hadronic center of mass frame) and 
integrated over its transverse momentum, $p_{h\perp}^2$. 
By defining the combination
\begin{equation}
M_q^h(x,z)=M_q^{h/P_1}(x,z)+M_q^{h/P_2}(x,z)\,,
\end{equation}
a straightforward generalization of eq.~(\ref{iDY}) leads to the parton model formula 
for the semi-inclusive Drell-Yan process:
\begin{equation}
\label{siDY}
\frac{d\sigma^{DY}(\tau)}{dQ^2 dz}=\frac{4\pi\alpha^2}{9 S Q^2}  
\int_{\tau}^{1-z} \frac{dx_1}{x_1} \int_{\frac{\tau}{x_1}}^{1} \frac{dx_2}{x_2}\sum_q e_q^2
\Big[ M_q^{h}(x_1,z) f_{\bar{q}}(x_2) +(x_1 \leftrightarrow x_2) \Big]
\,\delta\Big( 1-\frac{\tau}{x_1 x_2}\Big)\,. 
\end{equation}
\begin{figure}[t]
\begin{center}
\includegraphics[width=8cm,height=4cm,angle=0]{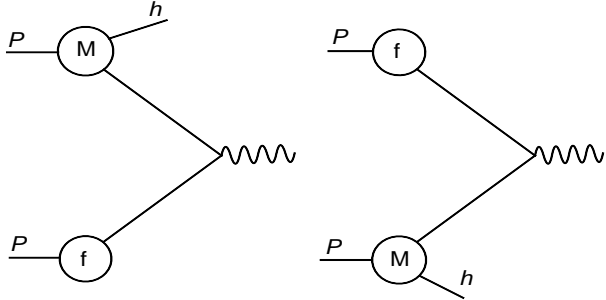}
\caption{\small A pictorial rapresentation of eq.~(\ref{siDY}). }
\label{Mf}
\end{center}
\end{figure}
\noindent
According to momentum conservation, the convolution integrals in eq.~(\ref{siDY}) must satisfy 
the constraints $1-z\ge x_1 x_2 \ge \tau$ and 
\begin{eqnarray}
x_1+z \le 1 \, \, \makebox{in} \,\, \mathcal{R}_{T_1} \,\,\makebox{and}\,\, 
x_2+z \le 1 \,\,\makebox{in} \,\,\mathcal{R}_{T_2}\,, \nonumber\\
x_2+z \le 1 \, \, \makebox{in} \,\, \mathcal{R}_{T_1} \,\,\makebox{and}\,\, 
x_1+z \le 1 \,\,\makebox{in} \,\,\mathcal{R}_{T_2} \,,\nonumber
\end{eqnarray} 
for the first and the second terms in square brackets of eq.~(\ref{siDY}) respectively.
The number of terms appearing in eq.~(\ref{siDY}) is twice 
the number appearing in the inclusive case since each fracture functions selects 
its own fragmentation region.  
Eq.~(\ref{siDY}) represents for the moment only a factorization \textsl{conjecture} for the process we are considering and 
is sketched in Fig.~(\ref{Mf}).
Our purpose is to show that eq.~(\ref{siDY}) does survive to the inclusion of radiative corrections and that the
factorization of collinear singularities is possible when  
the proper subtraction terms for bare fracture functions are added. The present
QCD-based calculation deals with the standard soft gluon 
exchange between active partons but it is blind to soft gluon exchange between spectators 
whose effects are taken into account in Refs.~\cite{soft_fact,CSS,Bodwin} and  
could spoil the factorization, as suggested in Ref.~\cite{Fact_M_soft}.
In the presence of such factorization breaking effects, as emerged in recent 
phenomenological analysis, 
diffractive parton distribution extracted from HERA data
do overestimate diffractive cross-sections measured at Tevatron by an order of magnitude~\cite{CDF}.  
Fracture functions appearing in eq.~(\ref{siDY}) therefore 
can not be related to the ones extracted from Semi-Inclusive DIS data. In a strong factorization breaking 
scenario the factorized form $M\otimes f$ itself could in principle be questioned, since 
the exchanges of low momemuntum gluons cannot be uniquely absorbed neither in the definition of 
fracture nor of parton distribution functions so that they pertain to the reaction as a whole.
The formalism we use can not indeed penetrate inside the details of soft factorization, however 
it may constitute a quantitative next-to-leading order guideline for estimating the magnitude of its breaking 
when passing from expected non-factorizing phase space region at low $p_{h\perp}^2$ to expected factorizing ones at high $p_{h\perp}^2$.
The possible identification of an intermediate scale or range of scales at which this transition 
may occur would constitute an important matter of inspection insight in the dynamics of the factorization mechanism.  
\vspace{0.7cm}

\noindent{\large \bf Evaluation of NLO corrections}\\

\noindent
As outlined above, we have adopted the normalization and conventions as of Ref.~\cite{DYNLO}. 
The cross-sections for the leading order partonic sub-process $q(p_1)+\bar{q}(p_2)\rightarrow \gamma^* $ 
is defined by:
\begin{equation}
\label{LO}
\frac{d\hat{\sigma}_{q\bar{q}}^{(0)}}{dQ^2}=\delta(1-w)\frac{1-\epsilon}{2N_C}\,,
\end{equation}
with $N_C$ the number of colours and $w=Q^2/s$ with $s=(p_1+p_2)^2$ 
the partonic center of mass energy. The space-time dimension is $n=4-2\epsilon$. 
Eq.~(\ref{LO}) is implicitely contained in parton model formula, eq.~(\ref{iDY}) and 
eq.~(\ref{siDY}).
Moving to next order in perturbation theory requires the evaluation of the 
real emission diagrams 
$q+\bar{q}\rightarrow \gamma^* + g$ and $q+g\rightarrow \gamma^* + q$  
along with virtual corrections to the Born amplitude. The results for the matrix 
elements squared, 
averaged over initial state colours and spins and 
summed over final ones, as well as the virtual contribution, can be found in Ref.~\cite{DYNLO}. 
Both the real and virtual contributions depend on the renormalization 
scale $\mu_r^2$.
Moreover, the cross-section dependence on the variables associated to the produced hadron 
is entirely contained in fracture functions for both Born and virtual contributions. 

So far no new ingredients have been added to a standard NLO perturbative calculations, 
the major subtleties coming from the two-particle final state 
of real emissions diagrams.
In order not to obscure the renormalization procedure we restrict ourselves   
to the discussion of the $q\bar{q}$ channel which indeed contains all the essential part  
of the calculations.  
In the following we will demonstrate that, at order $\mathcal{O}(\alpha_s)$,  
we can organize radiative corrections in such way that 
uncancelled (collinear) divergencies can be reabsorbed into the "bare" distributions 
appearing in eq.~(\ref{siDY}), in analogy with the semi-inclusive DIS calculation 
as performed in Ref.~\cite{Graudenz}.

When the final state hadron is observed in  
$\mathcal{R}_{T_1}$ or $\mathcal{R}_{T_2}$ (for istance in hard single diffractive 
events in hadronic collisions), we assume that it has been 
produced non-perturbatively from fracture functions. All the perturbative 
real radiation thus must be integrated over and virtual corrections added.
The corrections to the semi-inclusive Drell-Yan process in the target fragmentation regions 
turn out to be the same as in the inclusive case. 
The two body phase space can be easily integrated in the partonic center of mass frame.
It can be shown by explicit calculations that the double poles in the $\epsilon$-expanded results cancel 
in the sum of real and virtual term. This latter feature is peculiar of the $q\bar{q}$ channel.
For the singular contributions to the cross-section in the target regions we obtain 
\begin{eqnarray}
\label{siDY_target_sing}
\frac{d\sigma^{DY}_t (\tau)}{dQ^2 dz}=\frac{4\pi\alpha^2}{9 S Q^2}
\int_{\tau}^{1-z} \frac{dx_1}{x_1} \int_{\frac{\tau}{x_1}}^{1} \frac{dx_2}{x_2} 
\sum_q e_q^2 \Big[ M_q^h(x_1,z) f_{\bar{q}}(x_2) +(x_1 \leftrightarrow x_2)\Big]\cdot\nonumber\\
\cdot\Bigg[ \delta(1-w)
-\frac{2}{\epsilon}\frac{\alpha_s(\mu_r^2)}{2\pi}
P_{qq}(w) \Big( \frac{4 \pi \mu_r^2}{Q^2} \Big)^{\epsilon} 
\frac{\Gamma(1-\epsilon)}{\Gamma(1-2\epsilon)}\Bigg]\,,%\;\;\;\;&&
\end{eqnarray} 
with $w=\tau/x_1 x_2$ and $P_{ij}(w)$ the Altarelli-Parisi splitting function. 
\begin{figure}[t]
\begin{center}
\includegraphics[width=11cm,height=4cm,angle=0]{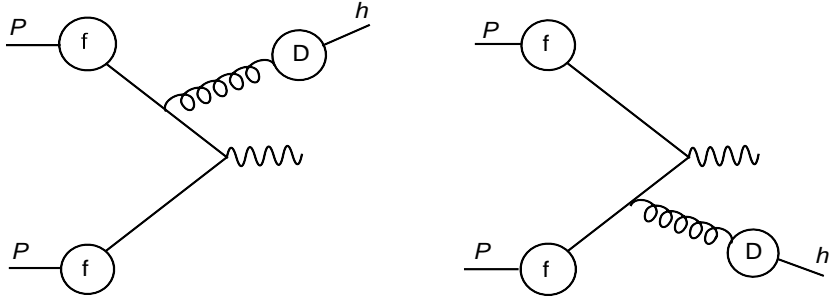}
\caption{\small Hadron production in the central region}
\label{ffd}
\end{center}
\end{figure}
Next we consider the production of the observed hadron $h$ by 
the fragmentation of a, real, final state parton (\textsl{i.e.} a gluon) 
in the partonic sub-process. 
This $\mathcal{O}(\alpha_s)$ production mechanism
is sketched in Fig.~(\ref{ffd}).
We address this phase space region as "central", where the hadron 
$h$ is allowed to be produced at high $p_{h\perp}^2$. 
In the collinear limit the "central" region collapses to the target 
fragmentation one. 
In general such correction is expected of the form~\cite{ACGG}
\begin{equation}
\label{form}
\frac{d\sigma^{DY}_c(\tau)}{dQ^2 dz}=\frac{4\pi\alpha^2}{9 S Q^2}
\sum_q e_q^2 \int \frac{d\rho}{\rho}\int \frac{dx_1}{x_1} \int \frac{dx_2}{x_2} 
f_q(x_1) f_{\bar{q}}(x_2) D_g^h(z/\rho)
\frac{d\hat{\sigma}_{q\bar{q}}^{g}}{dx_1 dx_2 d\rho}\,. 
\end{equation}
The variable $\rho=2E_k/\sqrt{S}$ is the partonic 
analogue of $z$, being $E_k$ the energy of 
the outgoing gluon and $d\hat{\sigma}_{q\bar{q}}^{g}$ the differential 
partonic cross-sections in the considered channel. 
The phase space integrations are performed in the hadronic rather then 
in the partonic center of mass frame.
The final state gluon depends on its fractional energy 
$\rho$ and on an angular variable $y=(1-\cos \theta_{cm})/2$, 
being $\theta_{cm}$ the relative angle between $k$ abd $P_1$.
These variables are however not independent and constrained
by the formula
\begin{equation}
\label{rho}
\rho(y)=\frac{2(x_1 x_2-\tau)}{x_1+x_2+(2y-1)(x_1-x_2)}\,.
\end{equation} 
In order to keep compact the expressions it is useful to rewrite 
the $\rho$ convolution as a $y$ integral, with 
$\rho$ depending on $y$ via eq.~(\ref{rho}).
The \textsl{central} contribution to the cross-sections therefore reads
\begin{equation}
\label{form2}
\frac{d\sigma^{DY}_c(\tau)}{dQ^2 dz}=\frac{4\pi\alpha^2}{9 S Q^2}
\sum_q e_q^2 \int_0^1 dy \int_{r_1(\tau,z;y)}^{1} \frac{dx_1}{x_1}
\int_{r_2(\tau,z;y)}^{1} \frac{dx_2}{x_2} f_q(x_1) f_{\bar{q}}(x_2) D_g^h(z/\rho)
\frac{d\hat{\sigma}_{q\bar{q}}^{g}}{dx_1 dx_2 dy} \frac{1}{\rho} \,. 
\end{equation}
The integration limits $r_1$ and $r_2$ 
are obtained by imposing momentum conservation and the condition $\rho \ge z$ in order 
to guarantee that the energy of the parent parton is greater
than the energy of the observed hadron:
\begin{equation}
r_1(\tau,z;y)=\frac{\tau+z(1-y)}{1-zy}\,,\;\;\;\;\;
r_2(\tau,z;y)=\frac{\tau+x_1 z y}{x_1-z(1-y)}\,.
\end{equation}
The emitted gluon in this region it is thus not allowed to be soft
since it is required to produce the observed hadron.
Once the invariants $t$ and $u$ appearing in the relevant matrix element
squared are specified in this frame, we perform 
a standard $\epsilon$-expansion of the result  
in two disjoint singular limits, \textsl{i.e.} for $y\rightarrow 0,1$.
Retaining only $\mathcal{O}(\epsilon^{-1})$ in the expansions we get the singular contributions 
to the cross-sections in this region of phase space:
\begin{eqnarray}
\label{central_sing}
\frac{d\sigma^{DY}_c(\tau)}{dQ^2 dz}&=&\frac{4\pi\alpha^2}{9S Q^2}
\int_{\tau+z}^{1}\frac{dx_1}{x_1} \int_{\tau/(x_1-z)}^{1} \frac{dx_2}{x_2} 
f_{q}(x_1) \, f_{\bar{q}}(x_2) \, D_g^h\Big(\frac{z x_2}{x_1 x_2 -\tau} \Big) 
\frac{\alpha_s(\mu_r^2)}{2\pi} \cdot \nonumber\\ 
&&\;\;\;\;\;\;\;\;\;\;\;\;\;\;\;\;\;\;\;\;\;\;\;\;\cdot \Big(-\frac{1}{\epsilon}\Big)
\frac{\Gamma(1-\epsilon)}{\Gamma(1-2\epsilon)} \Big( \frac{4 \pi \mu_r^2}{Q^2} \Big)^{\epsilon}
\hat{P}_{(g)q\leftarrow q}\,\Big(\frac{\tau}{x_1 x_2}\Big)\frac{x_2}{x_1 x_2-\tau}\nonumber\\
&&+\frac{4\pi\alpha^2}{9S Q^2}
\int_{\tau/(1-z)}^{1}\frac{dx_1}{x_1} \int_{z+\tau/x_1}^{1} \frac{dx_2}{x_2} 
f_{q}(x_1) \, f_{\bar{q}}(x_2) \, D_g^h\Big(\frac{z x_1}{x_1 x_2 -\tau} \Big)
\frac{\alpha_s(\mu_r^2)}{2\pi}\cdot \nonumber\\ 
&&\;\;\;\;\;\;\;\;\;\;\;\;\;\;\;\;\;\;\;\;\;\;\;\;\cdot  \Big(-\frac{1}{\epsilon}\Big)
\frac{\Gamma(1-\epsilon)}{\Gamma(1-2\epsilon)} \Big( \frac{4 \pi \mu_r^2}{Q^2} \Big)^{\epsilon}
\hat{P}_{(g)\bar{q}\leftarrow \bar{q}}\,\Big(\frac{\tau}{x_1 x_2}\Big)\frac{x_1}{x_1 x_2-\tau}\,.
\end{eqnarray}
The first term in eq.~(\ref{central_sing}) is singular for $y\rightarrow 0$ 
while the second  for $y\rightarrow 1$. In the previous expression 
do appear unregularized Altarelli-Parisi splitting functions,
$\hat{P}_{(g)\bar{q}\leftarrow \bar{q}}(w)$, since no 
interference with virtual contribution is present. 
Eqs.~(\ref{siDY_target_sing}) and eq.~(\ref{central_sing}) thus 
represents all the singular collinear contributions to the semi-inclusive Drell-Yan cross-sections. 
If we were considering the inclusive Drell-Yan case, the only 
singularities would be the ones shown in eq.~(\ref{siDY_target_sing}). The  
subtraction of singular term in the partonic cross-sections would be performed 
by lumping the divergence into the bare parton distributions $f$.
In the $\overline{\textrm{MS}}$ subtraction scheme it reads: 
\begin{equation}
\label{f_renorm}
f_i(\xi)=\int_{\xi}^{1}\frac{du}{u} \Big[ \delta_{ij}\delta(1-u) +\frac{1}{\epsilon}\frac{\alpha_s(\mu_r^2)}{2\pi}
\frac{\Gamma(1-\epsilon)}{\Gamma(1-2\epsilon)}
\Big( \frac{4 \pi \mu_r^2}{\mu^2} \Big)^{\epsilon} P_{ij}(u) \Big] f_j\Big(\frac{\xi}{u},\mu^2\Big)\,.
\end{equation}
In the previous equation $\mu_r^2$ and  $\mu^2$ are respectively the renormalization and 
factorization scale. Since any observable built using eq.~(\ref{f_renorm}) can not depend 
on $\mu^2$, the derivative with respect to $\ln \mu^2$ gives standard QCD evolution
equations~\cite{DGLAP}.
In the semi-inclusive case however there are additional singularities, eq.~(\ref{central_sing}),
and the subtraction of eq.~(\ref{f_renorm})
would not be sufficient to render the cross-sections infrared finite. 
Fracture functions, however, have been shown to have a more complex 
evolution equations with respect to the one of $f$. They contain an inhomogeneous term
which accounts for hadron production in the target fragmentation region of the projectile
by the showering of initial state radiation~\cite{Trentadue_Veneziano}.
The analogous of eq.~(\ref{f_renorm}) for bare fracture functions is obtained in the context of 
semi-inclusive Deep Inelastic Scattering~\cite{Graudenz}:
\begin{eqnarray}
\label{M_renorm}
M_i^h(\xi,\zeta)=\int_{\xi/(1-\zeta)}^{1}\frac{du}{u} \Big[ \delta_{ij}\delta(1-u) 
+\frac{1}{\epsilon}\frac{\alpha_s(\mu_r^2)}{2\pi}
\frac{\Gamma(1-\epsilon)}{\Gamma(1-2\epsilon)}
\Big( \frac{4 \pi \mu_r^2}{\mu^2} \Big)^{\epsilon} P_{ij}(u) \Big] 
M_j^h\Big(\frac{\xi}{u},\zeta,\mu^2\Big)\,+\nonumber\\
+\int_\xi^{\xi/(\xi+\zeta)} \frac{du}{u} \frac{1}{1-u}\frac{u}{\xi}
\frac{1}{\epsilon}\frac{\alpha_s(\mu_r^2)}{2\pi}
\frac{\Gamma(1-\epsilon)}{\Gamma(1-2\epsilon)}
\Big( \frac{4 \pi \mu_r^2}{\mu^2} \Big)^{\epsilon}
\hat{P}_{(k)i\leftarrow j}(u)\,f_j\Big(\frac{\xi}{u}\Big)\, D_k^h\Big( \frac{\zeta u}{\xi(1-u)}\Big)\,.
\end{eqnarray}
The first term on r.h.s of eq.~(\ref{M_renorm}) has the same subtraction structure as 
for parton distribution, eq.~(\ref{f_renorm}). The singularity is due do collinear 
radiation accompaining the active parton, while the hadron in the final state 
is non perturbatively produced by the fracture functions itself. 
In the second term of eq.~(\ref{M_renorm}) instead, the singularity is due to 
the observed hadron being collinear to the incoming hadron and generated by the 
perturbative fragmentation of the emitted collinear parton. 

At this point we insert eq.~(\ref{f_renorm}) and eq.~(\ref{M_renorm})
in the parton model result, eq.~(\ref{siDY}).
After some algebra and integral manipulation is easy but lenghtly to show 
that eq.~(\ref{f_renorm}) and the homogeneous term of eq.~(\ref{M_renorm})
produce exactly the pole term in eq.~(\ref{siDY_target_sing})
with the opposite sign. The inhomogeneous term 
in eq.~(\ref{M_renorm}) instead reproduce the singular term in eq.~(\ref{central_sing}). 
All the singularities therefore cancel and we are left with

\begin{eqnarray}
\label{siDY_finite_qq}
\frac{d\sigma^{DY}(\tau)}{dQ^2 dz}&=&\frac{4\pi\alpha^2}{9 S Q^2} \sum_q e_q^2 
\int_{\tau}^{1-z} \frac{dx_1}{x_1} \int_{\frac{\tau}{x_1}}^{1} \frac{dx_2}{x_2}
\cdot\Big[ M_q^{h}(x_1,z,\mu_F^2) f_{\bar{q}}(x_2,\mu_F^2) +(x_1 \leftrightarrow x_2) \Big]\cdot\nonumber\\
&&\;\;\;\;\;\;\;\;\;\;\;\;\;\;\;\;\;\;\;\;\;\;\;\;\;\;\;\;\;\;\;
\;\;\;\;\;\;\;\;\;\;\;\cdot\Bigg[ \delta\Big( 1-\frac{\tau}{x_1 x_2}\Big)+\frac{\alpha_s(Q^2)}{2\pi}
C_{qq}\Bigg( \frac{\tau}{x_1 x_2},\frac{\mu_F^2}{Q^2}\Bigg) \Bigg]+\nonumber\\
&&+\frac{4\pi\alpha^2}{9 S Q^2} \sum_q e_q^2 \int_0^1 dy \int_{r_1}^{1} \frac{dx_1}{x_1}
\int_{r_2}^{1} \frac{dx_2}{x_2} \Big[f_{q}(x_1,\mu_F^2) \, f_{\bar{q}}(x_2,\mu_F^2)+
(x_1 \leftrightarrow x_2)\Big] \cdot\nonumber\\ 
&&\;\;\;\;\;\;\;\;\;\;\;\;\;\;\;\;\;\;\;\;\;\;\;\;\;\;\;\;\;\;\;\;\;\;\;\;
\;\;\;\;\;\;\;\;\;\;\;\cdot D_g^h\Big(\frac{z}{\rho},Q^2\Big) \frac{\alpha_s(Q^2)}{2\pi} K_{q\bar{q}}^g
\Bigg(z,y,\frac{\tau}{x_1 x_2},\frac{\mu_F^2}{Q^2}\Bigg) \,.
\end{eqnarray}
The function  $C_{q\bar{q}}$ and $K_{q\bar{q}}^g$  are infrared finite
and depend explicitely on the factorization scale $\mu_F^2$. 
All bare distributions are replaced by renormalized as indicated by the 
explicit factorization scale dependence.
In particular $C_{q\bar{q}}$ is the same as in the inclusive Drell-Yan case
whereas  $K_{q\bar{q}}^g$ is specific of the semi-inclusive process. 
Its explicit form along with the coefficient for the gluon initiated channel will 
be reported in a separate paper, as well as a more detailed description of the calculation.
In eq.~(\ref{siDY_finite_qq}) we may set $\mu_F^2=Q^2$ in order to remove potentially large
logarithmic corrections of the type $\ln(\mu_F^2/Q^2)$ from the coefficient functions and resum
them by using the appropriate evolution equations for $f$ and $M$. 
\vspace{0.7cm}

\noindent{\large \bf Conclusions}\\

\noindent
We have shown, with a fixed $\mathcal{O}(\alpha_s)$ perturbative QCD calculation, 
that the partonic cross-sections for the semi-inclusive Drell-Yan process do factorize at the collinear 
level. We therefore implicitely confirm the general widespread idea indicating that
soft exchanges between active and spectators partons~\cite{CFS} as responsible for factorization breaking  in semi-inclusive reactions. 

We would like to conclude by listing a few remarks and proposals.
As opposed to full inclusive observables, semi-inclusive ones in hadronic collisions
are affected by soft gluon exchange contributions and could therefore act as factorization analyzers at  phenomenological level. 
In the central region, high $p_{h\perp}^2$ hadron production should follow the pattern predicted by perturbative QCD. 
When the detected hadron is, instead, a low $p_{h\perp}^2$ proton, diffractive processes may occur~\cite{UA8} and the partonic structure of the color-singlet exchanged object may be studied~\cite{IS}. 
In these particular cases a non universality of diffractive parton distributions, as taken from diffractive DIS
and hadronic collisions, was suggested in Ref.~\cite{BS}. This was experimentally reported in Ref.~\cite{CDF}.
It would be interesting to establish with a comparison with the data whether a factorization breaking shows up only in a diffractive kinematic regime or if it manifests itself also in processes with a gapless final state containing a single hadron as well in the target fragmentation region.
For the same reason it would be interesting also to study, within the proposed approach, 
light mesons production which is sensitive to the soft, high multiplicity, fragmentation process.
\newpage
The present work can be also generalized to double hadron production. 
The evaluation of a double  hadron production cross-section needs a full $\mathcal{O}(\alpha_s^2)$ QCD calculation. 
However, an approximate result could be obtained  if one considers two hadron at low $p_{h\perp}^2$ 
observed in opposite fragmentation regions with respect to the incoming hadrons. 
In this case higher order corrections for this process should be the same as for inclusive 
Drell-Yan process, when the proper kinematics is taken into account.   
In Ref.~\cite{M_isr} we have suggested an analogous formula, in leading logarithmic
approximation, for double-inclusive Drell-Yan production which also includes the additional dependence on the invariant momentum transfer at the proton's vertex $t_1$ and $t_2$.

Finally we are thinking to a generalization of the present approach 
to include gluon initiated hard processes~\cite{dijet}
whose relevance in diffractive Higgs production was first 
suggested in Ref.~\cite{Graudenz_Veneziano}.


\begin{thebibliography}{99}
\bibitem{DY}S.~D.~Drell, T.~Yan ~\textsl{Phys.~Rev.~Lett.~} \textbf{25}, 316 (1970);
Erratum-ibid.~\textbf{25}, 902 (1970);
\bibitem{soft_fact}W.~W.~Lindsay, D~.A.~Ross, C.~T.~Sachrajda, ~\textsl{Phys.~Lett.~}  \textbf{B117}, 105 (1982);
\textsl{Nucl.~Phys.~} \textbf{B214}, 61  (1983);
\textsl{ Nucl.~Phys.~} \textbf{B222}, 189 (1983);\\
W.~W.~Lindsay, ~\textsl{Nucl.~Phys.~}  \textbf{B231}, 397 (1984);
\bibitem{CSS} J.~C.~Collins, D.~E.~Soper, G.~Sterman, \textsl{Phys.~Lett.~} \textbf{B134}, 263 (1984); 
\textsl{Nucl.~Phys.~} \textbf{B261}, 104 (1985); \textsl{Nucl.~Phys.~} \textbf{B308}, 833 (1988);
\bibitem{Bodwin} G.~ T.~ Bodwin, \textsl{Phys.~Rev.~} \textbf{D31}, 2616 (1985); Erratum-ibid.
\textbf{D34}, 3932 (1986);
\bibitem{DeGrand} T.~A.~DeGrand, H.~I.~Miettinen, ~\textsl{Phys.~Rev.~Lett.~} \textbf{40}, 612 (1978);
\bibitem{CMS_TOTEM_TDR}M.~Albrow et al., CERN-LHCC-2006-039, CERN-LHCC-G-124, CERN-CMS-NOTE-2007-002;
\bibitem{joint} ATLAS, CMS, TOTEM Collaborations, L.~Fano et al, 
in *Hamburg 2007, Blois07, Forward physics and QCD*, 207-214;
\bibitem{DeTar} C.~E.~DeTar, S.~D.~Ellis, P.~V.~Landshoff,~\textsl{Nucl.~Phys.~} \textbf{B87}, 176 (1975);\\
J.~L.~Cardy, G.~A.~Winbow,~\textsl{Phys.~Lett.~} \textbf{B52}, 95 (1974);
\bibitem{STAR} STAR Collaboration, J.~Adams et al., ~\textsl{Phys.~Rev.~Lett.~} 97, 152302 (2006);
\bibitem{CFS} J.~C.~Collins, L.~Frankfurt, M.~Strikman,~\textsl{Phys.Lett.} \textbf{B307}, 161 (1993); 
\bibitem{BS}A.~Berera, D.~E.~Soper,~\textsl{Phys.~Rev.~} \textbf{D50}, 4328 (1994);
\bibitem{Fact_M_soft}J.C.~Collins,~\textsl{Phys.~Rev.~} \textbf{D57}, 3051 (1998);
\bibitem{DYNLO}G.~Altarelli, R.~K.~Ellis, G.~Martinelli,~\textsl{Nucl.~Phys.~} \textbf{B157}, 461 (1979);
\bibitem{Trentadue_Veneziano} L.~Trentadue, G.~Veneziano,~\textsl{Phys.~Lett.~} \textbf{B323}, 201 (1994);
\bibitem{Fact_M_coll} M.~Grazzini, L.~Trentadue, G.~Veneziano,~\textsl{Nucl.~Phys.~} 
\textbf{B519}, 394 (1998);
\bibitem{Graudenz} D.~Graudenz,~\textsl{Nucl.~Phys.~} \textbf{B432}, 351 (1994);
\bibitem{DeFSassot}D.~de~Florian, R.~Sassot,~\textsl{Phys.~Rev.~} \textbf{D58},054003 (1998);
\bibitem{DPDF} H1 Coll. A.~Aktas et al.,~\textsl{JHEP} \textbf{0710}, 042 (2007);\\
H1 Coll., A.~Aktas et al., ~\textsl{Eur.~Phys.~J.~} \textbf{C48}, 715 (2006);\\
ZEUS Coll., S.~Chekanov et al.,~\textsl{ Eur.~Phys.~J.~} \textbf{C38}, 43 (2004);
\bibitem{CDF} CDF Collaboration, ~\textsl{Phys.~Rev.~Lett.~} \textbf{84}, 5043, (2000);
\bibitem{ACGG}F.~Aversa, P.~Chiappetta, M.~ Greco, J.~P~.~ Guillet, ~\textsl{Nucl.~Phys.~} 
\textbf{B327}, 105 (1989);
\bibitem{DGLAP}L.N.~Lipatov,~\textsl{Sov.~J.~Nucl.~Phys.~}  \textbf{20}, 95 (1975);\\
V.N.~Gribov and L.N.~Lipatov,~\textsl{Sov.~J.~Nucl.~Phys.~}  \textbf{15}, 438 (1972);\\
G.~Altarelli and G.~Parisi,~\textsl{Nucl.~Phys.~} \textbf{B126}, 298 (1977);\\
Yu.L.~Dokshitzer \textsl{Sov.~Phys.~JETP~} \textbf{46}, 641 (1977);
\bibitem{UA8} UA8 Collaboration, ~\textsl{Phys.~Lett.~} \textbf{B297}, 41 (1992);
\bibitem{IS}G.~Ingelman, P.~E.~ Schlein, ~\textsl{Phys.~Lett.~} \textbf{B152}, 256 (1985);
\bibitem{M_isr}F.~A.~Ceccopieri, L.~Trentadue, ~\textsl{Phys.~Lett.~} \textbf{B655}, 15 (2007);
\bibitem{dijet}S.~Chekanov {\it et al.}  [ZEUS Collaboration],
  %``Dijet production in diffractive deep inelastic scattering at HERA,''
  Eur.\ Phys.\ J.\  C {\bf 52}, 813 (2007);\\
  A.~Aktas {\it et al.}  [H1 Collaboration],
  %``Dijet Cross Sections and Parton Densities in Diffractive DIS at HERA,''
  JHEP {\bf 0710}, 042 (2007);\\
    A.~A.~Affolder {\it et al.}  [CDF Collaboration],
 Phys.\ Rev.\ Lett.\  {\bf 88}, 151802 (2002);
\bibitem{Graudenz_Veneziano}D.~Graudenz, G.~Veneziano, ~\textsl{Phys.~Lett.~} \textbf{B365}, 302 (1996); 
\end{thebibliography}
\end{document}